\documentclass[nofootinbib, aps, prd, a4paper, 10pt, superscriptaddress, eqsecnum, showkeys]{revtex4-2}
\usepackage[a4paper, top=2cm, bottom=2cm, left=2.5cm, right=2.5cm]{geometry}

\usepackage{amsmath}
\usepackage{amsfonts}
\usepackage{amsthm}
\usepackage{bm}
\usepackage{mathrsfs}
\usepackage{braket}

\usepackage{graphicx}
\usepackage{booktabs}

\usepackage{xcolor}
\usepackage[pdfusetitle]{hyperref}
\hypersetup{colorlinks = true, allcolors = blue}

\usepackage{orcidlink}

\begin{document}

\title{Analytic $k$-essence Inflation}
\author{V.K. Oikonomou\orcidlink{0000-0003-0125-4160}}
\email{voikonomou@gapps.auth.gr} \affiliation{Department of
Physics, Aristotle University of Thessaloniki, Thessaloniki 54124,
Greece}\affiliation{Center for Theoretical Physics, Khazar
University, 41 Mehseti Str., Baku, AZ-1096, Azerbaijan}

\begin{abstract}
In this work we develop a fully analytic $k$-essence inflation
solution which is compatible with the observational constraints on
inflation, without the need for the slow-roll assumption being
imposed dynamically in the field equations. The analytic solution
is obtained for a kinetic scalar sector of the form $K(X)=
\frac{X^n}{X_s^{n-1}}$ and for the scalar field satisfying a
tracking condition $\dot{\phi}^2=\gamma H^{-m}$. For this class of
theories, the sound speed of the primordial curvature
perturbations is constant, so no gradient instabilities occur in
the theory. The Hubble rate, the scale factor and the scalar field
can be analytically expressed as functions of the cosmic time, and
the observational indices of inflation depend solely on the
parameters $n$ and $m$. The resulting evolution is inflationary
and it develops a pressure singularity a few $e$-foldings after
the end of inflation. However, using qualitative and quantitative
numerical arguments, we show that the pressure singularity is
classically avoided when the matter fluids of radiation and cold
dark matter are taken into account.
\end{abstract}

%PACS numbers: 04.50.Kd, 95.36.+x, 98.80.-k, 98.80.Cq

\maketitle

\section{Introduction}

General Relativity (GR) is undoubtedly a successful theory at an
astrophysical level, at least for the moment, however at large
scales it faces serious challenges. Specifically, the latest
observational data coming from the Dark Energy Spectroscopic
Instrument (DESI) \cite{DESI:2024mwx, DESI:2025zgx} indicated
quite interesting new physics that cannot be harbored in standard
GR contexts, since the dark energy seems to have a phantom
transition at late times. There is a considerable stream of
articles trying to model such a transition, but also the Hubble
tension, see for example \cite{Cortes:2024lgw, Colgain:2024xqj,
Giare:2024smz, Shlivko:2024llw,Odintsov:2024woi, Dai:2020rfo,
He:2020zns, Nakai:2020oit, DiValentino:2020naf,
Agrawal:2019dlm,Ye:2020btb,Desmond:2019ygn, Hogas:2023pjz,
OColgain:2018czj, Krishnan:2020obg, Colgain:2019joh, Lee:2022cyh,
Krishnan:2021dyb, Ye:2021iwa, Ye:2022afu, Verde:2019ivm,
Menci:2024rbq, Adil:2023ara, Reeves:2022aoi, Ferlito:2022mok,
Vagnozzi:2021quy, DiValentino:2020evt, Sabogal:2024yha,
CosmoVerseNetwork:2025alb, Odintsov:2025kyw, Odintsov:2025jfq,
Kessler:2025kju, Nojiri:2025low, Nojiri:2025uew,
Odintsov:2026fyc,Han:2024qbw,You:2024hit,He:2024hmc,Han:2024qbw,You:2024hit,He:2024hmc,
Tang:2024gtq,Wolf:2025acj,Wolf:2025jed,Wolf:2024stt,Li:2026xaz,Li:2025owk,OdintsovSGStens:2021,Vagnozzi:2021tjv,DiValentino:2025sru}.
These tensions, combined with the fact that the recent ACT data
for inflation \cite{ACT:2025fju,ACT:2025tim}, indicated that most
of GR-related inflationary models are in tension with the data at
least $2.5\sigma$, points out that gravity primordially and at
large scales, cannot be fully described in a consistent way by GR
and it needs modifications. Indeed, with regards to the ACT data,
quite many models in the context of modified gravity can be
compatible with the ACT data, see for example Refs.
\cite{Kallosh:2025rni,Gao:2025onc,Liu:2025qca,Yogesh:2025wak,Yi:2025dms,Peng:2025bws,Yin:2025rrs,Byrnes:2025kit,
Wolf:2025ecy,Aoki:2025wld,Gao:2025viy,Zahoor:2025nuq,Ferreira:2025lrd,Mohammadi:2025gbu,Choudhury:2025vso,
Odintsov:2025wai,Q:2025ycf,Zhu:2025twm,Kouniatalis:2025orn,Hai:2025wvs,Dioguardi:2025vci,Yuennan:2025kde,
Kuralkar:2025zxr,Kuralkar:2025hoz,Modak:2025bjv,
Aoki:2025ywt,Ahghari:2025hfy,McDonough:2025lzo,Chakraborty:2025wqn,NooriGashti:2025gug,Yuennan:2025mlg,
Deb:2025gtk,Afshar:2025ndm,Ellis:2025zrf,Iacconi:2025odq,Yuennan:2025tyx,Wang:2025cpp,Qiu:2025uot,Wang:2025dbj,
Asaka:2015vza,Oikonomou:2025htz,Choudhury:2025hnu,Singh:2025uyr,Kim:2025dyi,Peng:2026ofs,Odintsov:2026dss,Oikonomou:2026mvp}.
Also an interesting approach on how the Hubble tension may affect
the inflationary predictions was given in \cite{Sabogal:2026qvy}.

There is however a GR-compatible theory which is quite promising
in alleviating the tensions, the $k$-essence theory
\cite{Noh:2001ia,Hwang:2005hb,Hwang:2002fp,Kaiser:2013sna,ArmendarizPicon:1999rj,Chiba:1999ka,ArmendarizPicon:2000dh,Matsumoto:2010uv,ArmendarizPicon:2000ah,Chiba:2002mw,Malquarti:2003nn,
Malquarti:2003hn,Chimento:2003zf,Chimento:2003ta,Scherrer:2004au,Aguirregabiria:2004te,ArmendarizPicon:2005nz,Abramo:2005be,
Rendall:2005fv,Bruneton:2006gf,dePutter:2007ny,Babichev:2007dw,Deffayet:2011gz,Kan:2018odq,Li:2012vta,Unnikrishnan:2012zu,Brax:2014wla,Joshi:2024nez,Yang:2023rjt,Herrera:2023mas,Papanikolaou:2022did,Pareek:2021lxz,Lola:2020lvk,Dutta:2018xcz,
Mishra:2018tki,HosseiniMansoori:2024pdq,Rezazadeh:2014fwa}. In the
context of $k$-essence theory a unified description of dark energy
and dark matter can be achieved \cite{Mishra:2018tki} and also the
tensions of the $\Lambda$-Cold-Dark-Matter model can be alleviated
\cite{HosseiniMansoori:2024pdq}. Also, recently it was shown that
$k$-essence theory can realize a phantom-to-quintessence
transition \cite{Nojiri:2026uvn}. The appealing feature of
$k$-essence theory is that it is a relative simple theory and it
is still GR-compatible, and does not constitute an actual
modification of GR, like for example standard modified gravity
models like $F(R)$ gravity or even Einstein-Gauss-Bonnet theory,
see Refs. \cite{reviews1,reviews2,reviews3,reviews4} for reviews.
Essentially $k$-essence is the farthest border of GR-compatible
cosmology. In this work, we aim to provide a fully analytical
$k$-essence inflationary model which is compatible with
inflationary observational data, without the need of the slow-roll
assumption being imposed dynamically in the field equations. To
our knowledge this is a fully analytical $k$-essence model in the
literature that is compatible with inflationary observations. The
analytical solution is obtained under the assumption that the
scalar field of the theory obeys a tracking evolution of the form
$\dot{\phi}^2=\gamma H^{-m}$. This tracking condition is known to
also provide analytic solutions to minimally coupled single scalar
field models \cite{Oikonomou:2026mvp}, and it also has a physical
underlying motivation in terms of early Universe Wilsonian quantum
field theory \cite{Oikonomou:2026jdg}. The tracking condition
itself is used in a variant form in $k$-mouflage gravity to
provide a viable late-time behavior \cite{Brax:2014wla} with the
$k$-essence part kinetic term being of the form $K(X) \sim
X^{1+1/m}$, where
$X=\frac{1}{2}\partial^{\mu}\phi\partial_{\mu}\phi$. In addition,
a variant form of the tracking condition has been used in
covariant Galileon theories \cite{DeFelice:2010aj,Bellini:2012qn},
for dark energy purposes too. Specifically, in Refs.
\cite{DeFelice:2010aj,Bellini:2012qn} the tracking condition was
used in the form $\dot{\phi}^2=\gamma H^{-2}$, so in our notation
$m=2$. Thus in this work we shall demonstrate that viable fully
analytic $k$-essence inflationary solutions can be obtained if the
tracking condition $\dot{\phi}^2=\gamma H^{-m}$ is utilized by the
theory, with the kinetic term being of the form $K(X)=
\frac{X^n}{X_s^{n-1}}$. For this class of models, the sound speed
of the primordial curvature perturbations is an $n$-dependent
constant, so no gradient instabilities occur. The model $K(X)=
\frac{X^n}{X_s^{n-1}}$ with general potentials was studied in
Refs. \cite{Unnikrishnan:2012zu,Li:2012vta,Papanikolaou:2022did},
but in these cases, the solutions were obtained using the
slow-roll assumption being imposed dynamically in the field
equations, the potentials are free to choose and the solutions
were not analytic. In our case, the potential is fixed by the
tracking condition and the solution is fully analytic.

The Hubble rate, the scale factor and the scalar field can be
found as functions of the cosmic time and we show that the
evolution is inflationary and compatible with the ACT data.
However, the theory in vacuum develops a pressure singularity a
few $e$-foldings after the end of the inflationary era.
Nevertheless, this singularity is classically avoided in the
presence of matter and radiation perfect fluids, which start
contributing to the cosmological equations as the end of the
inflationary era approaches. Using convincing qualitative
arguments but also a strong numerical analysis, we show both
qualitatively and quantitatively that the pressure singularity is
avoided and the Universe passes smoothly through it in the last
few $e$-folds of the inflationary regime. This feature is also a
characteristic of the analytic single scalar field theory
\cite{Oikonomou:2026mvp}.

In the following we shall assume that the metric is that of a
Friedmann-Robertson-Walker (FRW) with flat spatial section,
\begin{equation}
\label{metricfrw} ds^2 = - dt^2 + a(t)^2 \sum_{i=1,2,3}
\left(dx^i\right)^2\, ,
\end{equation}
with $a(t)$ denoting as usual the scale factor. Note that the
metric convention we use is different compared to Ref.
\cite{Unnikrishnan:2012zu,Li:2012vta}, which studies the same
$k$-essence model we will study here, but using the slow-roll
condition imposed in the field equations and without analytic
solutions. Caution is needed with the field equations when the
convention of the metric (\ref{metricfrw}) is used. We will use
the convention of Ref. \cite{Hwang:2005hb}.

\section{General $k$-essence Inflation Formalism and Analytic $k$-essence Inflationary Phenomenology}

The $k$-essence theories of inflation have the general action,
\begin{equation}
\label{mainactionB} \mathcal{S}=\int d^4x\sqrt{-g}\left[
\frac{1}{2}f(R,\phi,X) \right]\, ,
\end{equation}
with $X=\frac{1}{2}\partial^{\mu}\phi\partial_{\mu}\phi$. The
perturbations in these theories were developed in Refs.
\cite{Noh:2001ia,Hwang:2005hb,Hwang:2002fp,Kaiser:2013sna}, and in
this work we shall use the conventions and notation of Ref.
\cite{Hwang:2005hb}. For the action (\ref{mainactionB}) if we vary
it with respect to the metric and the scalar field we obtain the
following field equations for a FRW metric \cite{Hwang:2005hb},
\begin{equation}\label{euqationsofmotion1}
3H^2F=f_{,X}X+\frac{RF-f}{2}-3H\dot{F}\, ,
\end{equation}
\begin{equation}\label{euqationsofmotion2}
-2\dot{H}F=Xf_{,X}+\ddot{F}-H\dot{F}\, ,
\end{equation}
\begin{equation}\label{euqationsofmotion3}
\frac{1}{a^3}\left( a^3\dot{\phi}f_{,X}\right)^{.}+f_{,\phi}=0\, ,
\end{equation}
with the ``dot'' denoting differentiation with respect to the
cosmic time and the ``prime'' denoting differentiation with
respect to the scalar field. Moreover, $F$ stands for
$F=\frac{\partial f}{\partial R}$. The sound speed of the scalar
primordial curvature perturbations has the form,
\begin{equation}\label{wavespeed}
c_A^2=\frac{Xf_{,X}+\frac{3\dot{F}^2}{2F}}{Xf_{,X}+2X^2f_{,XX}+\frac{3\dot{F}^2}{2F}}\,
,
\end{equation}
with $f_{,X}=\frac{\partial f}{\partial X}$ and
$f_{,XX}=\frac{\partial^2 f}{\partial X^2}$. For these theories,
the perturbations are quantified in terms of the slow-roll
indices, defined as \cite{Hwang:2005hb},
\begin{align}\label{slowrollparameters}
& \epsilon_1=-\frac{\dot{H}}{H^2}\, , \,\,\,
\epsilon_2=\frac{\ddot{\phi}}{H\dot{\phi}}\,
,\,\,\,\epsilon_3=\frac{\dot{F}}{2H F}\, ,\,\,\,
\epsilon_4=\frac{\dot{E}}{2HE}\, ,
\end{align}
with $E$ being defined as,
\begin{equation}\label{functionE}
E=-\frac{F}{2X}\left(X f_{,X}+2X^2 f_{,XX}+\frac{3\dot{F}^2}{2F}
\right)\, .
\end{equation}
The resulting observational indices are expressed in terms of the
slow-roll indices. Specifically, the spectral index of the
primordial scalar curvature perturbations is \cite{Hwang:2005hb},
\begin{equation}\label{spectralindex}
n_s=1-4\epsilon_1-2\epsilon_2+2\epsilon_3-2\epsilon_4\, ,
\end{equation}
and the tensor-to-scalar ratio reads,
\begin{equation}\label{tensortoscalarration}
r=16 \left(\epsilon_1-\epsilon_3 \right) c_A\, ,
\end{equation}
and recall $c_A$ was defined in Eq. (\ref{wavespeed}) for the
general $f(R,X,\phi)$ theory. Now we specialize the function
$f(R,\phi,X)$ in the action (\ref{mainactionB}) to be of the form,
\begin{equation}\label{generalfunctionf}
f(R,\phi,X)=\frac{R}{\kappa^2}-2K(X)-2V(\phi)\, ,
\end{equation}
in order to obtain the single scalar field action by simply taking
$K(X)=X$. For this class of models, the sound speed
(\ref{wavespeed}) becomes particularly simple,
\begin{equation}\label{wavespeed1}
c_A^2=\frac{Xf_{,X}}{Xf_{,X}+2X^2f_{,XX}}\, .
\end{equation}
For the needs of this study, the exact $f(R,\phi,X)$ gravity will
be of the form,
\begin{equation}\label{exactactionf}
f(R,\phi,X)=\frac{R}{\kappa^2}-2\frac{X^n}{X_s^{n-1}}-2V(\phi)\, ,
\end{equation}
where $X_s$ is a dimensionful parameter with mass dimensions
$[X_s]=[X]$. These models were firstly developed in Ref.
\cite{Li:2012vta} and lead to a sound speed of the form,
\begin{equation}\label{wavespeedfinal}
c_A^2=\frac{1}{2n-1}\, ,
\end{equation}
so for $n>\frac{1}{2}$ the theory is gradient instabilities free.
\begin{figure}
\centering
\includegraphics[width=28pc]{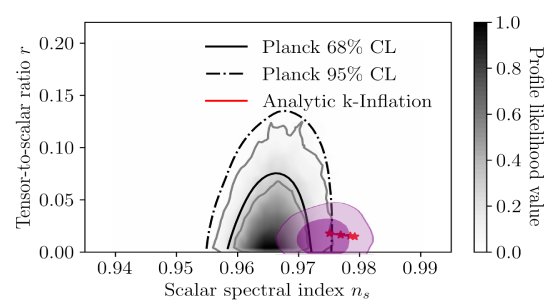}
\caption{The marginalized curves of the Planck 2018 data and the
$k$-essence analytic model, versus the ACT data, the Planck 2018
data, and the BICEP/Planck constraints on $r$ ratio, for the
$e$-foldings range $N\sim 50-60$ and $(m,n)=(2,3)$.}\label{plot1}
\end{figure}
Note that if we take $n=1$, the function (\ref{generalfunctionf})
results to the canonical and minimally coupled single scalar field
case. At this point we impose the tracking condition
\cite{Oikonomou:2026mvp},
\begin{equation}\label{trackingmain}
\dot{\phi}^2=\gamma H^{-m}\, ,
\end{equation}
where $\gamma$ is a dimensionful parameter and $m$ some positive
even integer. The field equations take an interesting form, with
the Friedmann equation becoming,
\begin{equation}\label{friedmannprefinal}
\frac{3
H^2}{\kappa^2}=-\frac{X^n}{X_s^{n-1}}\left(2n-1\right)+V(\phi)\, ,
\end{equation}
while the Raychaudhuri equation takes the form,
\begin{equation}\label{raychaudhuriprefinal}
\dot{H}=\frac{n\kappa^2 X^n}{X_s^{n-1}}\, ,
\end{equation}
and finally, the modified $k$-essence Klein-Gordon equation takes
the form,
\begin{equation}\label{scalarfieldprefinal}
3 n
H\dot{\phi}\frac{X^{n-1}}{X_s^{n-1}}+n\ddot{\phi}\frac{X^{n-1}}{X_s^{n-1}}+n(n-1)\dot{\phi}\frac{X^{n-2}}{X_s^{n-1}}\dot{X}+V'=0\,
.
\end{equation}
\begin{figure}
\centering
\includegraphics[width=28pc]{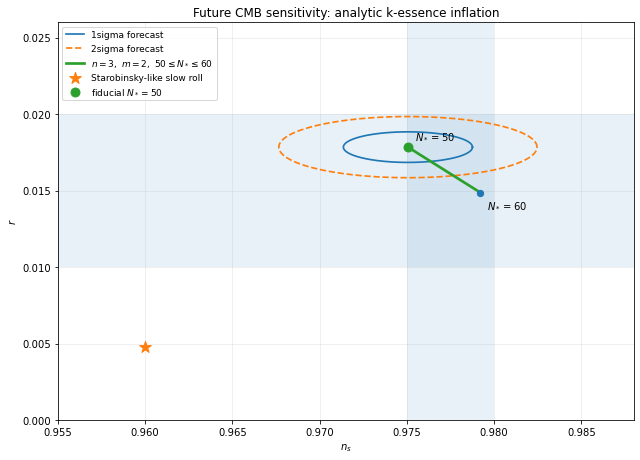}
\caption{Future-CMB sensitivity in the $n_s-r$ plane, for the
fiducial analytic k-essence inflationary model with $(n,m)=(3,2)$.
The solid and the dashed contours correspond to 1$\sigma$ and
2$\sigma$ regions, which are obtained by using the uncertainties
$\sigma(n_s)=0.0037$ and $\sigma(r)=0.001$. The contours are not
joint CMB likelihood contours and vanishing covariance between
$n_s$ and $r$ is assumed. The solid curve is obtained by varying
$N_*$ in the range $50-60$ for $(m,n)=(2,3)$.}\label{plot2a}
\end{figure}
It is easy to see that the above reduce to the canonical single
scalar field scenario for $n=1$. Since for the FRW metric
$X=-\frac{\dot{\phi}^2}{2}$, $n$ must be of the following form,
\begin{equation}\label{nvalues}
n=\frac{2\ell_1+1}{2\ell_2+1}\, ,
\end{equation}
with $\ell_1$ and $\ell_2$ being positive integers. Hence, the
Raychaudhuri equation reads,
\begin{equation}\label{raychaudhuriprefinal1}
\dot{H}=-\frac{n\kappa^2}{X_s^{n-1}}\left(\frac{\dot{\phi}^2}{2}\right)^n\,
.
\end{equation}
Using the tracking condition and for $n$ as in Eq.
(\ref{nvalues}), the Raychaudhuri equation takes a very simple
form,
\begin{equation}\label{raychaudhurifinal}
\dot{H}=-\mathcal{A}H^{-m n}\, ,
\end{equation}
with $\mathcal{A}=\frac{n\kappa^2}{2^{n+1}X_s^{n-1}}\gamma^n$.
Note that the differential equation (\ref{raychaudhurifinal}) is a
simple differential equation and admits an inflationary solution
for $n$ chosen as in Eq. (\ref{nvalues}). If one chooses
$n=\frac{2\ell_1}{2\ell_2+1}$, the differential equation becomes
$\dot{H}=\mathcal{A}H^{-m n}$ which does not describe conventional
inflation, but a super-inflationary phantom inflation regime,
which we will not consider here. By solving the Raychaudhuri
equation (\ref{raychaudhurifinal}) we obtain the following
analytic solution,
\begin{equation}\label{analytichubblerate}
H(t)=\big{(}m n+1\big{)}^{\frac{1}{m n+1}} \big{(}\omega
-\mathcal{A} t\big{)}^{\frac{1}{m n+1}}\, ,
\end{equation}
where $\omega$ is an integration constant. It is also easy to
obtain $\phi(t)$ analytically in this context, by solving
$\dot{\phi}^2=\gamma H^{-m}$, so we get,
\begin{equation}\label{phioftanalytic}
\phi(t)=\mathcal{C}_0+\frac{2 \sqrt{\gamma } (m n+1)^{1-\frac{m}{2
m n+2}} \big{(}\omega -\mathcal{A} t\big{)}^{1-\frac{m}{2 m
n+2}}}{\mathcal{A} \big{(}m (2 n-1)+2\big{)}}\, ,
\end{equation}
where $\mathcal{C}_0$ is an integration constant. From the
Friedmann equation (\ref{friedmannprefinal}) one specifies the
scalar potential which is a fixed function of the cosmic time
hereafter,
\begin{equation}\label{scalarpotentialfinal}
V\big{(}\phi(t)\big{)}=\frac{3
H^2}{\kappa^2}+\frac{X^n}{X_s^{n-1}}\left(2n-1\right)\, .
\end{equation}
Using the potential (\ref{scalarpotentialfinal}) and the
Raychaudhuri equation, the Klein-Gordon equation
(\ref{scalarfieldprefinal}) is trivially satisfied by using
$\dot{X}=-\dot{\phi}\ddot{\phi}$. The slow-roll indices also
become quite simple in this context, and specifically after some
simple algebra one gets,
\begin{equation}\label{epslionena}
\epsilon_1=\frac{\mathcal{A}}{H^{m
n+2}},\,\,\,\epsilon_2=\frac{m}{2}\epsilon_1,\,\,\,\epsilon_4=\frac{m(n-1)}{2}\,\epsilon_1\,
,
\end{equation}
and also the observational indices also acquire a very simple form
and can be expressed in terms of the first slow-roll index as
follows,
\begin{equation}\label{spectralindexasfunctionofepsilonena}
n_s=1-\left(m n+4
\right)\epsilon_1,\,\,\,r=\frac{16}{\sqrt{2n-1}}\,\epsilon_1\, .
\end{equation}
Combining the relations in
(\ref{spectralindexasfunctionofepsilonena}), one obtains a very
simple linear relation between the spectral index and the
tensor-to-scalar ratio,
\begin{equation}\label{simplelinearrelation}
n_s(r)=1-\frac{\left(m n+4 \right)\sqrt{2n-1}}{16}\,r\, ,
\end{equation}
see Ref. \cite{Oikonomou:2026vrw} for more general $n_s(r)$
relations in viable inflationary dynamics.
\begin{figure}
\centering
\includegraphics[width=25pc]{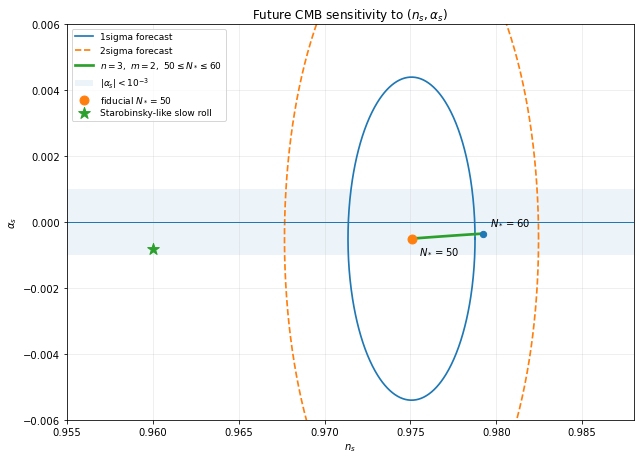}
\caption{Future-CMB sensitivity in the $a_s-n_s$ plane, for the
fiducial analytic k-essence inflationary model with $(n,m)=(3,2)$.
The solid and the dashed contours correspond to 1$\sigma$ and
2$\sigma$ regions, which are obtained by using the uncertainties
$\sigma(n_s)=0.0037$ and $\sigma(a_s)=0.0049$. The contours are
not joint CMB likelihood contours and vanishing covariance between
$n_s$ and $a_s$ is assumed. The solid curve is obtained by varying
$N_*$ in the range $50-60$ for $(m,n)=(2,3)$.}\label{plot2b}
\end{figure}
It is relatively simple to evaluate the observational indices of
inflation at first horizon crossing. Inflation ends when the
Hubble rate has the value $H_e$ determined by solving
$\epsilon_1(t_e)=\frac{\mathcal{A}}{H_e^{m n+2}}$ from Eq.
(\ref{epslionena}), so inflation ends when,
\begin{equation}\label{endofinflation}
H_e=\mathcal{A}^{\frac{1}{m n+2}}\, ,
\end{equation}
while using $N=\int_{t_i}^{t_e}H(t)\mathrm{d}t$ we find that
inflation starts when the Hubble rate was $H(t)=H_*$,
\begin{equation}\label{hubblerate}
 H_*=\Big{(}\mathcal{A}\Big{(}1+(m n+2)N_* \Big{)} \Big{)}^{\frac{1}{m
 n+2}}\, .
\end{equation}
Accordingly, evaluating the first slow-roll index at first horizon
crossing $H=H_*$ we have,
\begin{equation}\label{epsilon1athorizon}
\epsilon_1(H_*)=\frac{1}{1+(m n+2)N_*}\, ,
\end{equation}
therefore the spectral index of the primordial scalar
perturbations becomes a simple function of the parameters $m$ and
$n$ at first horizon crossing,
\begin{equation}\label{spectralindexfinalformhorizoncrossing}
n_s=1-\frac{m n+4}{1+(m n+2)N_*}\, ,
\end{equation}
while the tensor-to-scalar ratio is,
\begin{equation}\label{tensortoscalarfinalhorizoncrossing}
r=\frac{16}{\sqrt{2n-1}\,\Big{(}1+(m n+2)N_*\Big{)}}\, .
\end{equation}
Also the running of the spectral index is also easy to obtain in
this context, so using,
\begin{equation}\label{runningdef1}
a_s=\frac{\mathrm{d} n_s}{\mathrm{d} \ln k}=\frac{\mathrm{d}
n_s}{\mathrm{d}N}\frac{\mathrm{d}N}{\mathrm{d} \ln
k}=-\frac{1}{1-\epsilon_1}\frac{\mathrm{d}
n_{\mathcal{S}}}{\mathrm{d}N}\, ,
\end{equation}
we get,
\begin{equation}\label{runningofspectralindexfinal}
a_s=-\frac{m n+4}{m n N_*^2+2 N_*^2+N_*}\, .
\end{equation}
With the relations (\ref{spectralindexfinalformhorizoncrossing}),
(\ref{tensortoscalarfinalhorizoncrossing}) and
(\ref{runningofspectralindexfinal}) one can confront the
inflationary theory at hand with the observational constraints on
inflation, and we do this hereafter. The phenomenology is quite
simple because the constraints of inflation will pinpoint the
values of $n$ and $m$ that yield a viable phenomenology. Before
going to this phenomenology analysis, we must also constrain the
values $\gamma$ and $X_s$ must take, given $N_*$, $m$ and $n$. The
amplitude of the power spectrum of the primordial scalar
perturbations for the $k$-essence theory at hand, has the form,
\begin{equation}\label{amplitudeofpowerspectrum}
\mathcal{P}_R=\frac{\kappa^2H^2}{8\pi^2\epsilon_1c_A}\, ,
\end{equation}
so at horizon crossing we have,
\begin{equation}\label{powerspectrumathorizoncrossing}
\mathcal{P}_{R_*}=\frac{\kappa^2
\sqrt{2n-1}}{8\pi^2}\mathcal{A}^{\frac{2}{m n+1}}\Big{(} 1+(m
n+2)N_*\Big{)}^{\frac{m n+4}{m n+2}}\, ,
\end{equation}
so using the Planck 2018 constraint $\mathcal{P}_{R_*}\sim
2.19\times 10^{-9}$ \cite{Planck:2018jri}, one can directly
constrain the parameters $X_s$ and $\gamma$ using Eq.
(\ref{powerspectrumathorizoncrossing}). We shall take this into
account in the next section where we solve numerically the
cosmological system in the presence of perfect matter fluids.

Now let us focus on the viability of the inflationary model, and
it proves that the model is overall compatible with the ACT data,
which constraint the inflationary spectral index of the primordial
scalar perturbations as follows \cite{ACT:2025fju, ACT:2025tim},
\begin{align}
\label{act} n_{\mathcal{S}}=0.9743 \pm 0.0034\, .
\end{align}
In addition, the updated Planck/BICEP constraint on the
tensor-to-scalar ratio is \cite{BICEP:2021xfz},
\begin{align}
\label{planck} r<0.036\, .
\end{align}
Furthermore, the ACT constraint on the running of the spectral
index at 68$\%$ confidence is \cite{ACT:2025tim},
\begin{equation}\label{runningactconstraint}
\frac{\mathrm{d}n_{\mathcal{S}}}{\mathrm{d}\ln k}=0.0062 \pm
0.0052\, ,
\end{equation}
and the  95$\%$ confidence ACT+Planck+lensing+DESI (P-ACT-LB)
constraint on the running of the spectral index is,
\begin{equation}\label{95confidence}
-0.004\leq a_s\leq 0.016\, .
\end{equation}
Proceeding to the phenomenology, the analytic $k$-inflation is
overall compatible with the ACT constraints on the inflationary
indices, the viability can occur for a wide range of the
parameters $(m,n)$ for the $e$-foldings range $N_*\sim 50-60$. Let
us consider here a benchmark viability set which we will use
hereafter as a reference fiducial model. For $m=2$ and $n=3$ and
for $N_*=50$ one gets $n_s=0.975062$ and $r=0.0178439$ and
$a_s=-0.000498753$ and for $N_*=55$ one gets $n_s=0.977324$,
$r=0.0162254$ and $a_s=-0.00041238$, so both the sets of values
yield ACT-compatible results. In Fig. \ref{plot1} we present the
Planck 2018 likelihood curves and the ACT data curves versus the
$k$-inflation model, for $m=2$ and $n=3$ and $N_*=[50,60]$. As it
can be seen the model is well fitted in the ACT data. We can
further analyze the inflationary phenomenology by considering some
Cosmic Microwave Background (CMB) data and to try to see the
statistical significance of the viability of the model.
\begin{figure}
\centering
\includegraphics[width=25pc]{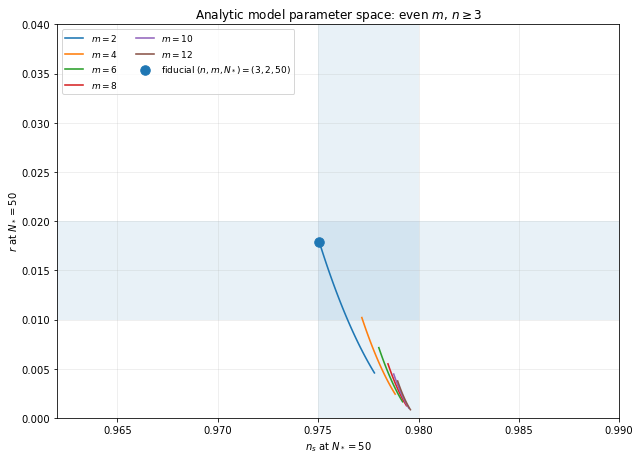}
\caption{The trajectories in the $(n_s,r)$-plane for  $N_*\sim 50$
and $3\leq n\leq 8$ and $m=2,4,...,12$.}\label{plot3}
\end{figure}
Let us take $N_*=50$ to be the fiducial point, which yields the
following predictions,
\begin{equation}
(n_s,r,\alpha_s)= \left( 0.975062, 0.017844, -4.988\times10^{-4}
\right) . \label{eq:fiducial_observables}
\end{equation}
Just for direct comparison, the predictions of the model for
$N_*=55$ are,
\begin{equation}
(n_s,r,\alpha_s)= \left( 0.977324, 0.016225, -4.123\times10^{-4}
\right),
\end{equation}
In the plots of Figs. \ref{plot2a} and \ref{plot2b} we present the
confrontation of the model with the CMB measurements. In Fig.
\ref{plot2a} we consider the $n_s-r$ plane and in Fig.
\ref{plot2b} we consider the $a_s-n_s$ plane, and we also added
the Starobinsky model for $N_*=50$ in Fig. \ref{plot2a}. For the
plots, we chose the CMB forecast errors as follows,
\begin{equation}
\sigma(n_s)=0.0037, \qquad \sigma(\alpha_s)=0.0049, \qquad
\sigma(r)=0.0010. \label{uncertanties}
\end{equation}
We need to note that the plots in Figs. \ref{plot2a} and
\ref{plot2b} are simple predictions relative to the given
sensitivities and these are not a detailed forecast or an actual
experimental likelihood. So the plots are not likelihood contours,
but these are generated by considering vanishing covariance
between the variables of the model. A detailed analysis would
require a detailed Fisher matrix analysis. The main characteristic
signature of the fiducial $k$-essence inflation model at $N_*=50$
is the following,
\begin{equation}
\alpha_s=-4.988\times10^{-4},
\end{equation}
and the forecast uncertainty is,
\begin{equation}
\sigma(\alpha_s)=4.9\times10^{-3},
\end{equation}
and simultaneously a blue-tilted spectral index is predicted,
\begin{equation}
n_s\simeq0.975\text{--}0.98, \qquad r\simeq0.015\text{--}0.018,
\label{eq:characteristic_signature}
\end{equation}
Also in Fig. \ref{plot3} we present the trajectories in the
$(n_s,r)$-plane by choosing $N_*=50$ and $3\leq n\leq 8$ and
$m=2,4,...,12$. As it can be seen, the model is well fitted in the
ACT data and also the model has a quite distinctive signature with
a blue-tilted scalar spectral index compared to the standard
slow-roll models.

Before closing, let us consider here an issue related with our
claim that the slow-roll era is not needed in this context. We
need to carefully put some context in this claim, because it is
rather unjustified. The ordinary condition for having a slow-roll
era is that $\dot{\phi}^2\ll V$ in standard slow-roll inflationary
theories. Also the condition $\epsilon_1\ll 1$ in our context does
not realize slow-roll regime but guarantees that inflation occurs
in the first place because $\epsilon_1$ defined in Eq.
(\ref{slowrollparameters}) is $\epsilon_1=-\frac{\dot{H}}{H^2}$,
so basically $\epsilon_1\ll 1$ is a realization of inflation, and
when $\epsilon_1\sim 1$ inflation ends. Thus in our context we did
not impose $\dot{\phi}^2\ll V$ by hand, however the condition
(\ref{trackingmain}) and subsequently Eq.
(\ref{raychaudhuriprefinal}) end up realizing the slow-roll
evolution as we now evince. Thus the tracking condition ends up to
be an actual slow-roll condition, without however being originally
intended for this purpose. Indeed, differentiating Eq.
(\ref{trackingmain}), we get
\begin{equation}\label{refrev1}
\frac{\ddot{\phi}}{H\dot{\phi}} =-\frac{m}{2}\frac{\dot
H}{H^2}=\epsilon_2 =\frac{m}{2}\epsilon_1\, ,
\end{equation}
as can also be seen in Eq. (\ref{epslionena}). Also, the same
applies for $\epsilon_4$, that is, it can be expressed in terms of
$\epsilon_1$ as can be seen in Eq. (\ref{epslionena}). Thus indeed
from the $\epsilon_2$ relation, it is shown that the scalar field
velocity varies very slowly during the inflationary regime in
which $\epsilon_1\ll 1$. In addition, we can compare the scalar
potential with the scalar field kinetic energy density. We have,
\begin{equation}\label{revref2}
V(H)=\frac{3H^2}{\kappa^2}
-\frac{(2n-1)\gamma^n}{2^nX_s^{n-1}}H^{-mn}\, ,
\end{equation}
and the positive kinetic contribution to the scalar energy density
is,
\begin{equation}\label{revref3}
\rho_{\rm kin} =\frac{(2n-1)\gamma^n}{2^nX_s^{n-1}}H^{-mn}\, .
\end{equation}
In the primordial era, where the Hubble rate is high, the
potential is dominated by its $3H^2/\kappa^2$ contribution, we
consequently obtain,
\begin{equation}\label{revref3}
\frac{\rho_{\rm kin}}{V} \simeq\frac{2n-1}{3n}\epsilon_1\ll 1\, .
\end{equation}
Thus the tracking condition has another effect, that of realizing
the slow-roll era, although we did not initially intend to use it
for this purpose. This is an important side effect which we needed
to present.

\section{Pressure Singularity in the Pure Scalar Theory and Classical Avoidance of the Pressure Singularity}

The Hubble rate of Eq. (\ref{analytichubblerate}) becomes zero at
\begin{equation}\label{singularitytime}
t_s=\frac{\omega}{\mathcal{A}}\, ,
\end{equation}
however the Universe at this time instance is singular, since
$\dot{H}$ becomes infinite. Thus according to the finite-time
singularity classification \cite{Nojiri:2005sx}, the singularity
is a pressure singularity (Type II singularity) because at
$t=t_s$, the scale factor is finite $a \to a_s$, the energy
density is also finite $\rho_\mathrm{eff} \to 0$ and only the
effective pressure diverges $\left|p_\mathrm{eff}\right| \to
\infty$. The end of inflation occurs at the time instance,
\begin{equation}\label{endofinflationtime}
t_f=t_s-\frac{\mathcal{A}^{-\frac{1}{2+m n}}}{m n+1}\, ,
\end{equation}
and by using $N_{post}=\int_{t_f}^{t_s}H(t)\mathrm{d}t$ we can
find how many $e$-foldings beyond the end of inflation the
singularity occurs. We easily find that $N_{post}=\frac{1}{m
n+2}$, so for a viable set of values for $m$ and $n$, for example
$m=2$ and $n=3$, we obtain $N_{post}= 0.125$. Thus the singularity
in the pure scalar theory occurs $N_{post}= 0.125$ after the end
of inflation. There is a caveat though in this line of thinking,
having to do with the fact that this scenario only occurs for the
pure scalar theory. However, the Universe is not void during
inflation, perfect matter fluids are present, specifically
radiation and cold dark matter perfect fluids. These redshift as,
\begin{equation}\label{eqrev10rad}
\rho_r\propto a^{-4}\, ,
\end{equation}
\begin{equation}\label{eqrev10matt}
\rho_m\propto a^{-3}\, ,
\end{equation}
so at the beginning of inflation these redshift as $\rho_r\sim
e^{-4 N_*}$ and $\rho_m\sim e^{-3 N_*}$, thus the matter fluid
contributions are subdominant at the beginning of inflation and
probably for the whole duration of it. But since the energy  of
the scalar field $\rho_{\phi}$ in pure scalar theory becomes zero
at the singularity $t=t_s$, this means that as the singularity is
approached, the matter fluids will start contributing to the total
energy density, and thus the Hubble rate will be modified.
Therefore, the total Hubble rate as the singularity is approached
is not given by Eq. (\ref{analytichubblerate}), but it will be
determined by the total scalar and matter fluids Friedmann
equation,
\begin{equation}\label{friedmannfull}
3\mathcal{H}^2=\kappa^2\left(\rho_{\phi}+\rho_r+\rho_m\right)\, .
\end{equation}
Hence, it is quite evident that the Hubble rate at the last stages
of inflation will be determined by the total system comprised by
the scalar field and the matter perfect fluids. This will occur
when $\rho_{\phi}\sim \rho_m+\rho_r$ in the Friedmann equation
(\ref{friedmannfull}) at which point the tracking condition will
be disrupted and thus the analytic Hubble rate
(\ref{analytichubblerate}) will no longer describe the evolution
of the Universe. The new Hubble rate will be the solution of the
total Friedmann equation (\ref{friedmannfull}). Apparently this is
a qualitative argument which is strong enough to motivate us to
perform a detailed numerical analysis in order to see if the
pressure singularity of the pure scalar system is indeed
classically avoided by the presence of the perfect matter fluids
of matter and radiation. So hereafter we test numerically whether
the singularity is eliminated by the inclusion of the perfect
matter fluids. Our approach is the following, we shall not impose
the constraint $\dot{\phi}^2=\gamma H^{-m}$  in the total
cosmological system, and instead we use the form of the potential
for the scalar field which was dictated by the constraint
$\dot{\phi}^2=\gamma H^{-m}$, given in Eq.
(\ref{scalarpotentialfinal}). For reading convenience, we quote
the form of the scalar potential here,
\begin{equation}\label{scalarpotentialfinalnewshow}
V\big{(}\phi(t)\big{)}=\frac{3
H^2}{\kappa^2}+\frac{X^n}{X_s^{n-1}}\left(2n-1\right)\, .
\end{equation}
Thus we will solve numerically the system and we aim to understand
when the analytical solution for the Hubble rate deviates from the
true solution for $H(t)$ determined by the scalar field and matter
fluids system.
\begin{figure}
\centering
\includegraphics[width=18pc]{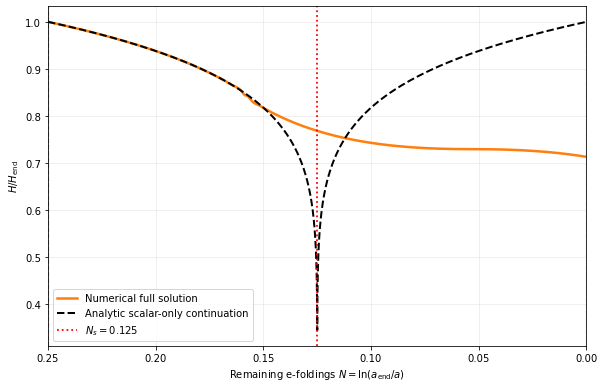}
\includegraphics[width=18pc]{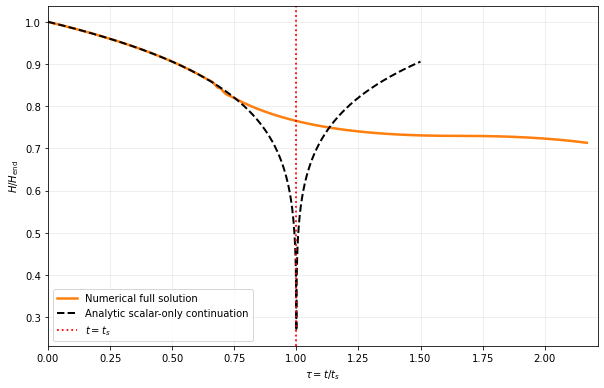}
\includegraphics[width=18pc]{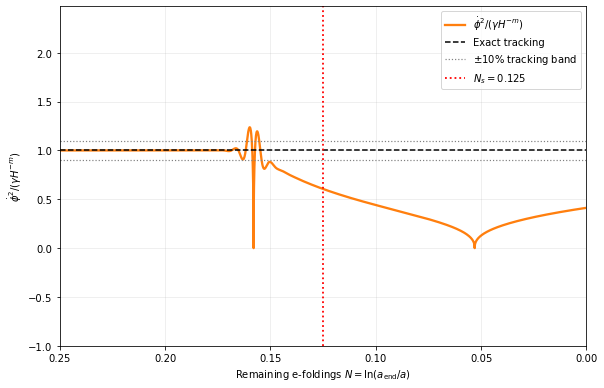}
\caption{The numerical Hubble rate in orange thick curve and the
analytical Hubble rate in dashed black, before and after the would
be finite-time singularity at $N_s$, as a function of the
$e$-foldings (upper left plot), the numerical Hubble rate in
orange thick curve and the analytical Hubble rate in dashed black,
before and after the would be finite-time singularity at $t=t_s$
as a function of $\tau=\frac{t}{t_s}$ (upper right plot) and the
diagnostic parameter $\mathcal{T}=\frac{\dot{\phi}^{2}}{\gamma
H^{-m}}$ as a function of the $e$-foldings number. Notice that the
sudden drop in the bottom plot is not singular, it is just a
sudden drop in the value of
$\mathcal{T}=\frac{\dot{\phi}^{2}}{\gamma H^{-m}}$.}\label{plot4}
\end{figure}
With our numerical method we will monitor when $\dot{\phi}^2$
deviates from $\gamma H^{-m}$. This numerical approach provides a
numerical separation between the analytic and numerically
constructed Hubble rates. Thus $\dot{\phi}^2$ and $H^{-m}$ evolve
dynamically as distinct quantities and are not constrained. Only
the scalar field potential is fixed to be the one in Eq.
(\ref{scalarpotentialfinalnewshow}) and this is a powerful
approach because it enables us to have full control on
implementing the tracking condition effects in the full
matter-scalar theory, while in parallel elevates the tracking
condition as something to be investigated if it holds true, not as
a condition imposed by hand in the system. For the numerical plots
and analysis we shall use the $e$-foldings number and also the
dimensionless cosmic time $\tau \equiv \frac{t}{t_s}$, where $t_s$
is the time instance that the scalar field only system reaches the
pressure singularity, with the corresponding $e$-foldings number
at the singularity is $N_s$. For the numerical technique, we use
the following numerical state vector,
\begin{equation}
    {\bf y}=
    \left(
    \frac{H}{H_{\rm end}},
    a,
    \phi,
    \frac{\dot{\phi}}{\dot{\phi}_{\rm end}},
    \frac{\rho_m}{\rho_{\rm crit,end}},
    \frac{\rho_r}{\rho_{\rm crit,end}}
    \right)\, ,
\end{equation}
and the evolution equations are basically, the continuity
equations of the perfect fluids, the Raychaudhuri equation and the
generalized Klein-Gordon equation,
\begin{equation}
    \dot H
    =
    -\frac{\kappa^2}{2}
    \left[
        \rho_\phi+P_\phi
        +\rho_m+\frac{4}{3}\rho_r
    \right]\, ,
\end{equation}
\begin{equation}
    \dot{\rho}_m=-3H\rho_m\, ,
    \qquad
    \dot{\rho}_r=-4H\rho_r\, .
\end{equation}
The system of differential equations is comprised by first order
differential equations and we integrate the system using the LSODA
python solver, with relative and absolute tolerances  $10^{-7}$
and $10^{-9}$. The integration stops when the $e$-foldings number
$N=0$ is reached or if the numerical Hubble rate changes sign.
Regarding the initial conditions, these are specified at the end
of inflation as follows,
\begin{equation}
    \dot{\phi}_{\rm end}
    =
    \sigma\sqrt{\gamma H_{\rm end}^{-m}}\, ,
\end{equation}
and $\phi_{\rm end}$ is the origin of the reconstructed scalar
field coordinate. Then $X_{\rm end}$ and $V_{\rm end}$ determine
the pressure and energy density of the scalar field. Also for the
matter fields we set,
\begin{equation}
    \Omega_{m,\rm end}=10^{-4}\, ,
    \qquad
    \Omega_{r,\rm end}=10^{-6}\, .
\end{equation}
The results of our numerical analysis are presented in Fig.
\ref{plot4}. In the upper left plot, we present the numerical
Hubble rate in orange thick curve and the analytical Hubble rate
in dashed black, before and after the would be finite-time
singularity at $N_s$, as a function of the $e$-foldings. In the
upper right we plot the numerical Hubble rate in orange thick
curve and the analytical Hubble rate in dashed black, before and
after the would be finite-time singularity at $t=t_s$. Also in the
bottom plot we present the diagnostic parameter
$\mathcal{T}=\frac{\dot{\phi}^{2}}{\gamma H^{-m}}$ as a function
of the $e$-foldings number. As it can be seen in the two upper
plots of Fig. \ref{plot4}, the numerical total solution and the
analytic solution remain identical until the very few last
$e$-foldings of inflation, at which point they deviate from each
other, with the numerical total Hubble rate passing smoothly
through the singularity. It is apparent that the numerical outcome
of the analysis is that the singularity is avoided in the total
scalar field and matter fluids system. Also in the bottom plot of
Fig. \ref{plot4} we can see that the tracking condition
$\dot{\phi}^2=\gamma H^{-m}$ begins to be violated at the same
$e$-foldings number that the analytic Hubble rate deviates from
the numerical Hubble rate. Notice that the sudden drop in the
bottom plot of Fig. \ref{plot4} is not singular, it is just a
sudden drop in the value of
$\mathcal{T}=\frac{\dot{\phi}^{2}}{\gamma H^{-m}}$. In conclusion,
the tracking condition $\dot{\phi}^2=\gamma H^{-m}$ holds for the
whole inflationary era, save the last $N\sim 0.05$ $e$-foldings,
thus $k$-essence viable inflation is realized even for the total
matter fluids scalar system, but more importantly near the last
0.05 $e$-foldings of inflation, the classical fluids start to
contribute to the evolution and they eliminate the singularity of
the scalar field system.

\section{Conclusions}

We developed a fully analytic inflationary solution for
$k$-essence theories, choosing the kinetic sector of the scalar
field to be of the form $K(X)= \frac{X^n}{X_s^{n-1}}$, and also
enforcing the scalar field to obey a tracking condition of the
form $\dot{\phi}^2=\gamma H^{-m}$. This class of $k$-essence
models leads to a constant sound wave speed for the primordial
scalar curvature perturbations of the form $c_A^2=\frac{1}{2n-1}$,
so no gradient instabilities occur for these theories. As we
showed, the tracking condition $\dot{\phi}^2=\gamma H^{-m}$
enabled us to solve analytically the Raychaudhuri equation and
thus obtain the Hubble rate of the theory as a function of the
cosmic time. The scalar field is also obtained as a function of
the cosmic time, thus the observational indices of inflation can
be evaluated fully analytically without the slow-roll assumption
being imposed dynamically in the field equations. As we showed,
the slow-roll indices can be expressed in terms of the first
slow-roll index of inflation and thus the resulting observational
indices depend solely on the parameters $m$ and $n$. Thus the
resulting inflationary theory is a two-parameter inflation theory,
which in fact can be compatible with the ACT observational data.
In addition, the analytic solution develops a pressure finite-time
singularity a few $e$-foldings after the end of inflation. This
singularity is developed however by the scalar field in vacuum,
and when classical matter fluids are included in the cosmological
evolution, the singularity is avoided. We show explicitly the
singularity avoidance by using qualitative and numerical
quantitative arguments. We described in detail the numerical
investigation we performed to show when the matter fluids start to
contribute significantly in the cosmological evolution and we
pinpointed numerically the time instance where the tracking
condition ceases to hold true, at which point the matter fluids
start to contribute significantly in the evolution. Thus we showed
in a solid way that the pressure singularity is classically
avoided and the Universe passes smoothly through the would be
pressure singularity in the presence of perfect matter fluids.
This is a fully analytic solution for inflationary $k$-essence
theories, which is obtained without the slow-roll assumption being
imposed dynamically in the field equations and it is also
compatible with the observational constraints on inflation.

\end{document}